\documentclass[aps,prd,preprint,floatfix,nobibnotes,nofootinbib]{revtex4-2}

\usepackage{amsmath}
\usepackage{amssymb}
\usepackage{graphicx}
\usepackage{float}
\usepackage{dsfont}
\usepackage{ulem}
\usepackage[colorlinks=true]{hyperref}

\newcommand{\be}{\begin{equation}}
\newcommand{\ee}{\end{equation}}
\newcommand{\ba}{\begin{eqnarray}} 
\newcommand{\ea}{\end{eqnarray}} 
\newcommand{\nn}{\nonumber}

\newcommand{\bea}{\begin{eqnarray}}
\newcommand{\eea}{\end{eqnarray}}

\usepackage[usenames, dvipsnames]{color}

\numberwithin{equation}{section}

\usepackage{setspace}

\begin{document} 

\title{Gauge invariant momentum broadening of hard probes in glasma}

\author{Margaret E. Carrington}
\affiliation{Department of Physics, Brandon University,
Brandon, Manitoba R7A 6A9, Canada}
\affiliation{Winnipeg Institute for Theoretical Physics, Winnipeg, Manitoba, Canada}

\author{Bryce T. Friesen}
\affiliation{Department of Physics, University of Toronto, Toronto, Ontario M5S 1A7, Canada}

\author{Stanis\l aw Mr\' owczy\' nski} 
\affiliation{National Centre for Nuclear Research, ul. Pasteura 7,  PL-02-093 Warsaw, Poland}

\date{June 25, 2026}

\begin{abstract}

We compute the transport coefficient $\hat q$ which quantifies the transverse momentum broadening of hard probes passing through the evolving glasma from the earliest stage of relativistic heavy-ion collisions. We use a proper-time expansion method which is designed to study the glasma at very early times. In our earlier calculations of $\hat q$ we used an approximation that greatly simplifies the complexity of the calculation but introduces a violation of gauge invariance. Based on these results we argued that the glasma plays an important role in jet quenching. In this paper we have used a gauge invariant formulation to calculate $\hat q$. The results for the momentum broadening coefficient are quantitatively very close to those of our previous simplified version of the calculation and confirm our earlier conclusion about the importance of the glasma contribution to jet quenching.

\end{abstract}

\maketitle
\newpage

%%%%%%%%%%%%%%%%%%%%%%%%%%%%%%%%%%%%%%%%%%%%%%%%%%%%%%%
\section{Introduction}
\label{sec-intro}
%%%%%%%%%%%%%%%%%%%%%%%%%%%%%%%%%%%%%%%%%%%%%%%%%%%%%%%

Hard probes are created through high momentum transfer processes at the earliest stage of a heavy-ion collision. These hard partons propagate through the system that is produced in the collision and lose a substantial fraction of their initial energy. This energy loss is responsible for the phenomenon of jet quenching which is treated as a signal of quark-gluon plasma, because only a deconfined phase of matter could cause a sufficient braking of hard probes. The history of over four decades of studies of jet quenching is told in the article \cite{Wang:2025lct}. The current status of experimental and theoretical research is summarized in the recent reviews \cite{Cunqueiro:2021wls,Cao:2024pxc,Mehtar-Tani:2025rty}. 
 
The matter produced in relativistic heavy-ion collisions equilibrates, at least partially, in less than 1~fm/$c$. The equilibrated quark-gluon plasma evolves hydrodynamically over a much longer timescale, for about 10 fm/$c$. For this reason it was expected that the main contribution to jet quenching comes from the equilibrium phase and for a long time the effect of the preceding nonequilibrium phases was entirely ignored. We note that two distinct pre-equilibrium phases should be identified: one shortly before the quark-gluon plasma reaches equilibrium when quasi-particles are already present in the plasma but their momentum distributions are not yet of equilibrium form, and the truly earliest phase which can be described in the framework of a Colour Glass Condensate (CGC) effective theory and is called glasma, see the reviews \cite{Iancu:2003xm,Gelis:2010nm,Gelis:2012ri}. The glasma phase consists of highly populated chromodynamic fields which can be treated as classical fields. 

It was suggested \cite{Mrowczynski:2017kso} that the short-lived glasma phase can contribute significantly to jet quenching due to its very high energy density which makes it opaque to hard probes. This suggestion was substantiated by our  analytic calculations performed using a proper time expansion method \cite{Carrington:2020sww,Carrington:2022bnv,Carrington:2021dvw,Carrington:2023nty} and in numerical simulations \cite{Ipp:2020mjc,Ipp:2020nfu,Avramescu:2023qvv,Avramescu:2024poa}, see also \cite{Boguslavski:2020tqz,Barata:2024xwy}.

Our previous calculations of the transport coefficient $\hat{q}$, which controls the radiative energy loss of a hard probe, suffered from an important deficiency. The coefficient $\hat{q}$ is determined by two-point correlators of chromodynamic glasma fields at different spacetime points. These correlators are gauge invariant when a Wilson link operator is inserted between the two fields in the correlation function (see Eq.~(\ref{tensor-X})). In our previous work we simplified the calculation by setting this link operator to unity. The motivation is that enormous technical difficulties are avoided when the proper time expansion of the vector potentials in the exponential link operator is not combined with the proper time expansion of the original field variables. We justified the procedure by showing that the expectation value of the link operator itself is about $0.84$ which is close to one \cite{Carrington:2021dvw}. The aim of this work is to calculate a gauge invariant result for the coefficient $\hat q$ and compare it with the results from the simpler calculations in our previous papers \cite{Carrington:2020sww,Carrington:2022bnv,Carrington:2021dvw,Carrington:2023nty} where the link operator was set to one. 

Throughout the paper we use the natural system of units with $c = \hbar = k_B =1$. The indices $\alpha, \beta = 1, 2, 3$ label the Cartesian spatial coordinates $(x,y,z)$ while the indices $i, j$ are reserved for transverse coordinates $(x, y)$. The generators $t^a$ of the SU($N_c$) group are defined through $[t^a,t^b] =  i f^{abc}t^c$ where $f^{abc}$ are the structure constants. The indices $a, b, c = 1, 2, \dots N_c^2 -1$ numerate colour components in the adjoint representation. The generators in the fundamental representation $t^a$ are normalized by ${\rm{Tr}}[t^a t^b] =\frac{1}{2} \delta^{ab}$. The generators in the adjoint representation are $[T^c]_{ab} = -i f^{abc}$ and ${\rm Tr}[T^a T^b] = N_c \delta^{ab}$. We use three systems of coordinates: Minkowski $(t, z, x, y)$, light-cone $(x^+,x^-, x, y)$ and Milne $(\tau, \eta, x, y)$, where $x^\pm = (t\pm z)/\sqrt{2}$, $\tau=\sqrt{t^2-z^2}$ and $\eta=\ln(x^+/x^-)/2$. Since we study chromodynamics only, the prefix `chromo' is neglected henceforth when referring to chromoelectric or chromomagnetic fields. We also consistently call the link operator (\ref{W-def}) a Wilson line.

%%%%%%%%%%%%%%%%%%%%%%%%%%%%%%%%%%%%%%%%%%%%%%%%%%%%%%%
\section{Formulation of the problem}
%%%%%%%%%%%%%%%%%%%%%%%%%%%%%%%%%%%%%%%%%%%%%%%%%%%%%%%

We calculate the coefficient $\hat{q}$ that determines the transverse momentum broadening of a hard probe as
\be
\label{qhat-X-T}
\hat q = \frac{2}{v} \Big(\delta^{\alpha\beta} - \frac{v^\alpha v^\beta}{v^2}\Big) X^{\alpha\beta}(\vec{v}) 
\ee
where $\vec v$ is the velocity of the probe. The tensor $X^{\alpha\beta}$ is defined as 
\be
\label{tensor-X}
X^{\alpha\beta}(\vec{v}) =
\frac{1}{2 N_c} \int_0^t dt' \big\langle F^i_a(t,\vec x) \, 
W_{ab}(t,\vec x\,|\,t',\vec y) \, F^j_b(t',\vec y) 
\big\rangle ,
\ee
where $\vec x- \vec y=\vec v (t-t')$, $\vec F_a(t,\vec x) = g\big(\vec E_a(t,\vec x)+\vec v\times \vec B_a(t,\vec x)\big)$ is the Lorentz colour force in the adjoint representation, $g$ is the coupling constant, and the Wilson line $W_{ab}(t,\vec x \,|\,t',\vec y)$ is the $(a,b)$ element of the matrix
\be
\label{W-def}
W(t,\vec x \,|\, t', \vec y) = {\cal P} \, 
{\rm exp}\Big(ig\int^{(t,\vec x)}_{(t',\vec y)} dz_\mu A^\mu_c(z) T^c \Big) .
\ee
The line integral follows the probe's trajectory from $\vec y$ at time $t'$ to $\vec{x}$ at $t$ and the symbol ${\cal P}$ denotes path ordering. In our calculation we will take the path to be a straight line between the two end points. 

The expression for the momentum broadening coefficient in Eqs.~(\ref{qhat-X-T}) and (\ref{tensor-X}) with \break $W = \mathds{1}$ was obtained in \cite{Mrowczynski:2017kso} where the transport of hard probes through glasma was described using a Fokker-Planck equation derived under the assumption that the hard probe interacts with a classical chromodynamic field. The applicability of the Fokker-Planck equation requires that the evolution of the fields is sufficiently slow that its characteristic time scale is much longer than the temporal correlation length of the field. In Appendix \ref{app-qhat-deriv} we show that the formula we use for $\hat q$ can actually be derived in a more general way, directly from the equation of motion of a hard probe in an external chromodynamic field. This means that our formula for $\hat q$ is not limited by the applicability of a transport equation and the limitations of the Fokker-Planck approach are irrelevant. We note that the insertion of the Wilson line in (\ref{qhat-X-T}) is motivated by the requirement of gauge invariance and not obtained from either derivation. 

For technical reasons that are explained in detail below, the presence of the Wilson line makes the calculation very difficult. In several previous papers \cite{Carrington:2022bnv,Carrington:2021dvw,Carrington:2023nty} we calculated $\hat q$ using a simplified approach in which we set $W = \mathds{1}$ and argued that this approximation gives qualitatively accurate results. In this paper we include the Wilson line and calculate the coefficient $\hat q$ using gauge invariant field correlators, and compare with the results of our previous calculations.

%%%%%%%%%%%%%%%%%%%%%%%%%%%%%%%%%%%%%%%%%%%%%%%%%%%%%%%
\section{Computational method}
\label{sec-method}
%%%%%%%%%%%%%%%%%%%%%%%%%%%%%%%%%%%%%%%%%%%%%%%%%%%%%%%

We summarize below the method to calculate the transport coefficient $\hat{q}$ of a hard probe in glasma using a proper time expansion, for more details see \cite{Carrington:2022bnv}.

%======================================================
\subsection{Glasma fields and proper time expansion}
\label{sec-fields}
%======================================================

We consider a collision of two heavy ions moving with the speed of light towards each other along the $z$ axis and colliding at $t=z=0$. The vector potential of the gluon field is described with the ansatz \cite{Kovner:1995ts} 
\ba
\nn
A^+(x) &=& \Theta(x^+)\Theta(x^-) x^+ \alpha(\tau,\vec x_\perp) ,
\\\label{ansatz}
A^-(x) &=& -\Theta(x^+)\Theta(x^-) x^- \alpha(\tau,\vec x_\perp) ,
\\ \nn
A^i(x) &=& \Theta(x^+)\Theta(x^-) \alpha_\perp^i(\tau,\vec x_\perp)
+\Theta(-x^+)\Theta(x^-) \beta_1^i(x^-,\vec x_\perp)
+\Theta(x^+)\Theta(-x^-) \beta_2^i(x^+,\vec x_\perp) ,
\ea
where the function $\beta_n^i(x^-,\vec x_\perp)$ with $n\in\{1,2\}$ represents the pre-collision potential of one of the two incoming ions, and the functions $\alpha(\tau,\vec x_\perp)$ and $\alpha_\perp^i(\tau,\vec x_\perp)$ are the post-collision potentials. In the forward light-cone the vector potential satisfies the sourceless Yang-Mills equations but the sources enter through boundary conditions that connect the pre-collision and post-collision potentials. The boundary conditions are
\ba
\label{cond1}
\alpha^{i}_\perp(0,\vec{x}_\perp) &=& \alpha^{i(0)}_\perp(\vec{x}_\perp) 
= \lim_{\text{w}\to 0}\left(\beta^i_1 (x^-,\vec{x}_\perp) + \beta^i_2(x^+,\vec{x}_\perp)\right) ,
\\
\label{cond2}
\alpha(0,\vec{x}_\perp) &=& \alpha^{(0)}(\vec{x}_\perp) 
= -\frac{ig}{2}\lim_{\text{w}\to 0}\;\big[\beta^i_1 (x^-,\vec{x}_\perp),\beta^i_2 (x^+,\vec{x}_\perp) \big] ,
\ea
where the notation $\lim_{\text{w}\to 0}$ indicates that the width of the sources across the light-cone is taken to zero, as the colliding nuclei are infinitely contracted.

To find solutions of the Yang-Mills equations that are valid for early post-collision times, the glasma potentials are expanded in the proper time $\tau$ as
\be
\label{alpha-exp}
\alpha^i_\perp(\tau, \vec{x}_\perp) =
\sum_{n=0}^\infty \tau^n \alpha^{i(n)}_{\perp} (\vec{x}_\perp), \qquad\qquad
\alpha(\tau, \vec{x}_\perp) = \sum_{n=0}^\infty \tau^n \alpha^{(n)} (\vec{x}_\perp). 
\ee
The  Yang-Mills  equations for the expanded glasma potentials are solved recursively so that higher order expansion coefficients are given in terms of the zeroth order coefficients, which are written in terms of the pre-collision potentials and their derivatives due to the boundary conditions (\ref{cond1}) and (\ref{cond2}).

%======================================================
\subsection{Two-point correlators}
\label{sec-corr}
%======================================================

The next step of the calculation is to use the Yang-Mills equations to express the pre-collision potentials through the colour charge distributions of the incoming ions. One then averages over a Gaussian distribution of colour charges within each nucleus. The average of a product of colour charges is written as a sum of products of the averages of all possible pairs, which is called Wick's theorem. We use the Glasma Graph Approximation (GGA) \cite{Lappi:2017skr} which means that we apply Wick's theorem not to colour charges but to gauge potentials. The average of a product of an odd number of pre-collision potentials vanish.

It is important to be able to test the accuracy of the GGA which greatly simplifies the calculations. This issue has been discussed in Sec.~5.2 of our paper \cite{Carrington:2020ssh}  where we obtained a quantitiative estimate of the reliability of the GGA in our calculations as follows. The pre-collision potentials generated by each of the incoming ions are pure gauge which means that their field strength tensor vanishes identically: $F_n^{ij}({\vec x}_\perp) = 0$ with $n\in\{1,2\}$. This means that an expectation value of any product of pre-collision potentials and their derivatives which includes $F_n^{ij}({\vec x}_\perp)$ vanishes. If we calculate the same product of pre-collision potentials by substituting the explicit form of the field strength tensor $F_n^{ij}=\partial^i\beta^j_n -\partial^j\beta_n^i-ig[\beta^i_n,\beta^j_n]$ and applying Wick's theorem to products of the potentials (instead of  products of colour charge densities), the result is non-zero. This non-zero result is a simple consequence of the fact that the GGA sets all products of an odd number of pre-collision potentials to zero (several examples are given in \cite{Carrington:2020ssh}).

The fact that these two calculations, which should be equivalent, give different results provides a way to test the reliability of the GGA. We computed several physical characteristics of glasma, in particular the ratios of the transverse and longitudinal pressure to energy density, $p_T/\varepsilon$ and $p_L/\varepsilon$, using the two different methods described above: first we set the pure gauge field strength tensor to zero before applying Wick's theorem, and second we write this field strength tensor in terms of potentials and then use Wick's theorem. The results for $p_T/\varepsilon$ and $p_L/\varepsilon$ calculated in these two different ways are shown in Fig.~2 of Ref.~\cite{Carrington:2020ssh}. Comparing the blue and red curves for the longitudinal pressure, the average difference and the largest difference are approximately 5\% and 17\%. The closeness of the curves representing the same quantity but computed differently is an indication of the accuracy of the GGA. 

The correlator of two pre-collision potentials from different ions is assumed to be zero because the potentials are not correlated to each other. All of the physical quantities we study are constructed from the correlators for two potentials from the same ion 
\ba
\label{field-corr}
\delta_{ab} B_n^{ij}(\vec{x}_\perp,\vec y_\perp) = \lim_{{\rm w}\to 0} \langle \beta^i_{na}(x^\mp,\vec{x}_\perp) \beta^j_{nb}(y^\mp,\vec y_\perp)\rangle ,
\ea
where $n \in \{1,2\}$ enumerates the two (right and left moving) ions and the upper/lower sign on the light-cone variables corresponds to the first/second ion. 

The correlator (\ref{field-corr}) can be calculated analytically and written as a function of the relative and average distances $\vec r = \vec x_\perp-\vec y_\perp$ and $\vec R = (\vec x_\perp + \vec y_\perp)/2$. In the original McLerran-Venugopalan (MV) formulation of the CGC approach, which is called the MV limit, the colliding nuclei are treated as infinite and uniform in the transverse plane, which means that the two-point correlator does not depend on $\vec R$. In the MV limit we can suppress the subscript $n$ since the two nuclei are identical. The MV limit gives another simplification in the momentum broadening calculation which is that we can choose the perpendicular component of the probe's velocity to lie along the $x$-axis, without loss of generality. Thus we write $\vec v_\perp = (v_\perp,0)$ and since $\vec r = \vec x_\perp-\vec y_\perp = \vec v_\perp t$ we have $\hat r \equiv \vec r/|\vec r|= (1,0)$. 

Since the original development of the CGC effective theory, many calculations have been done that go beyond the MV~limit. In the context of proper time expansions, a method to consider finite nuclei with realistic density profiles was introduced in \cite{Chen:2015wia} and further developed for both calculations of the energy-momentum tensor and associated observables \cite{Carrington:2021qvi}, and calculations of the momentum broadening and energy loss of a hard probe \cite{Carrington:2023nty}. The results show, unsurprisingly, that while the effect of nuclear structure is crucial for some observables it is essentially unimportant for others. For the momentum broadening coefficient the effects of nuclear structure give only small corrections relative to the MV limit \cite{Carrington:2023nty}. We therefore consider only infinite transverse nuclei which means that the two-point correlator depends only on $\vec r$  and is denoted as $B^{ij}(\vec{r})$.

The two-point correlator (\ref{field-corr}) in the MV~limit was derived for the first time in \cite{Jalilian-Marian:1996mkd} and can be written in the form \cite{Carrington:2022bnv} 
\be
\label{B-res2}
B^{ij}(\vec r)
= \delta^{ij}  C_1 (r) - \hat{r}^i \hat{r}^j C_2 (r) 
\ee
with the functions $C_1(r)$ and $C_2(r)$ given as
\ba
\label{C1-def}
C_1(r) &\equiv&  \frac{m^2 K_0 (mr)}{g^2 N_c  \big( mr K_1(mr) - 1\big) }
\bigg\{ \exp\bigg[\frac{g^4 N_c \mu \big( mr K_1(mr) - 1\big)  }{4\pi m^2 }\bigg] -1 \bigg\} \, ,
\\[2mm]
\label{C2-def}
C_2(r) &\equiv&  \frac{m^3 r \, K_1(mr) }{g^2 N_c  \big( mr K_1(mr) - 1\big) }
\bigg\{ \exp\bigg[\frac{g^4 N_c \mu \big( mr K_1(mr) - 1\big)  }{4\pi m^2 }\bigg] -1 \bigg\} \, ,
\ea
where $m = \Lambda_{\rm QCD}$ is an infrared cutoff and $K_0(z), ~K_1(z)$ are the Macdonald functions or modified Bessel functions of the second kind. 

We write derivatives of $B^{ij}(\vec{r})$ with respect to the components of $\vec r$ using subscripts, so that the subscript (10) indicates $-\partial/\partial r_x$ and (01) indicates $-\partial/\partial r_y$, where $r_x$ and $r_y$ are the $x$- and $y$-components of $\vec r$. The tensor indices are written as superscripts using an obvious notation. We list below the non-zero results for the two point correlator and its derivatives up to second order
\be
\label{deriv-res}
\begin{aligned}
B^{xx}(\vec r) &= C_1(r)-C_2(r) ,
\\
B^{yy}(\vec r) &= C_1(r) ,
\\
B^{xx}_{(0,2)}(\vec r) &= \frac{2 C_2(r)}{r^2} + \frac{C_1'(r)}{r}-\frac{C_2'(r)}{r} ,
\\
B^{xx}_{(1,0)}(\vec r) &= -{\rm sgn}(r_x) \left(C_1'(r) - C_2'(r)\right) ,
\\
B^{xx}_{(2,0)}(\vec r) &=  C_1''(r) - C_2''(r) ,
\\
B^{xy}_{(0,1)}(\vec r) &= B^{yx}_{(0,1)}(\vec r) = {\rm sgn}(r_y) \, \frac{C_2(r)}{r} ,
\\
B^{xy}_{(1,1)}(\vec r) &= B^{yx}_{(1,1)}(\vec r) = {\rm sgn}(r_x) \, {\rm sgn}(r_y) 
\Big( \frac{C_2(r)}{r^2} - \frac{C_2'(r)}{r} \Big),
\\
B^{yy}_{(0,2)}(\vec r) &= \frac{C_1'(r)}{r} - \frac{2 C_2(r)}{r^2} ,
\\
B^{yy}_{(1,0)}(\vec r) &= - {\rm sgn}(r_x)\, C_1'(r) ,
\\
B^{yy}_{(2,0)}(\vec r) &= C_1''(r) , 
\end{aligned}
\ee
where the vector $\vec r$ is chosen along the axis $x$ and $\hat r=(1,0)$. We note that some care is needed when calculating these derivatives, as some correlators vanish but their derivatives do not. One must first compute the derivatives and then set $\vec r=(r_x,0)$. 

%======================================================
\subsection{Regularization}
\label{sec-regular}
%======================================================

It is easy to check that the function $C_1(r)$ given by Eq.~(\ref{C1-def}) diverges logarithmically as $r \to 0$. The reason is that the CGC approach we are working with is classical and not valid at arbitrarily small distances. In the calculation of the momentum broadening coefficient these ultraviolet divergences appear in two ways. The first can be seen in Eq.~(\ref{tensor-X}) where the spatial arguments of the two fields become equal at the lower limit of the integral. There are also divergent terms in the integrand itself. This happens because when the field correlators in the tensor (\ref{tensor-X}) are written in terms of chromodynamic potentials, one-point correlators  $B^{ij}(\vec{x}_\perp,\vec x_\perp) = B^{ij}(\vec{r} = 0)$ appear due to the terms in the Yang-Mills equations that are non-linear in the potentials.

In our previous works \cite{Carrington:2022bnv,Carrington:2021dvw,Carrington:2023nty} we considered two different regularization methods and showed that results were largely independent of the method used. Neither of these methods  is well suited for the current calculation for a reason explained at the end of Sec.~\ref{sec-qhat}. In this paper we use a slightly different regularization. First we explain our previous method. We defined the regularized correlator
\be
\label{regularization1}
B^{ij}(\vec x_\perp,\vec y_\perp) = B^{ij}_{\rm reg}(\vec r) \equiv \Theta(r_s - r) \, B^{ij}(r_s) 
+ \Theta(r - r_s) \, B^{ij}(\vec r) , 
\ee
where 
\be
B^{ij}(r_s) \equiv \delta^{ij}C_1(r_s) - \hat r^i\hat r^j \, C_2(r_s) \,.
\ee
In this way the singular part of the correlator is cut off at a distance $r_s=Q_s^{-1}$. Derivatives of correlators were regulated analogously using
\be
\partial^k B^{ij}(\vec r)\big|_{\rm reg} = \Theta(r_s - r) \, \partial^k B^{ij}(\vec r) \big|_{r=r_s} 
+ \Theta(r - r_s) \, \partial^k B^{ij}(\vec r)\,.
\label{regularization1-der}
\ee
All one-point correlators were treated as the corresponding two-point correlator in the limit that the two points approach each other. For example, we used
\be
\label{one-point-def}
B^{ij}(\vec x_\perp,\vec x_\perp) = \lim_{\vec y_\perp\to\;\vec x_\perp}B^{ij}_{\rm reg}(\vec x_\perp,\vec y_\perp) 
= B^{ij}(r_s)
\ee
and similarly for all one-point correlators involving derivatives.

In this work we use the same method to regulate two-point correlators (Eqs.~(\ref{regularization1}) and (\ref{regularization1-der})). The change is that for one-point functions we use (\ref{one-point-def}) only for correlators that are not differentiated, and one-point correlators that involve derivatives are simply set to zero. Since the vector $\vec r$ has a non-zero component only in the $x$-direction the components of the one-point correlator are
\ba
B^{xx}(\vec x_\perp,\vec x_\perp) &=& C_1(r_s) - C_2(r_s), 
\\
B^{yy}(\vec x_\perp,\vec x_\perp) &=& C_1(r_s),
\\
B^{xy}(\vec x_\perp,\vec x_\perp) &=& B^{yx} (\vec x_\perp,\vec x_\perp) = 0 
\ea
(see the first two equations in (\ref{deriv-res})).  At the end of Sec.~\ref{sec-qhat} we explain why this change in the regularization method is needed and compare the new method with the one used in our previous calculations. We show that the results for the momentum broadening coefficient from the two methods differ only slightly.

%======================================================
\subsection{Probe velocity}
%======================================================

The momentum broadening coefficient depends strongly on the direction of the velocity of the hard probe, as discussed in detail in \cite{Carrington:2022bnv}. Experimentally the most important configuration is the one studied at both RHIC and the LHC, which is the probe moving perpendicularly to the beam direction. This is also the configuration in which our results are most accurate and reliable because when the probe's velocity is perpendicular to the beam axis the coefficient $\hat q$ as a function of time reaches its maximum before the proper time expansion breaks down \cite{Carrington:2022bnv}. For these reasons we consider in this work only the case where $\vec v$ is exactly perpendicular to the beam, which means $v^z=0$. We also take $v=1$.

%%%%%%%%%%%%%%%%%%%%%%%%%%%%%%%%%%%%%%%%%%%%%%%%%%%%%%%
\section{Average Wilson line}
\label{sec-Wilson-line}
%%%%%%%%%%%%%%%%%%%%%%%%%%%%%%%%%%%%%%%%%%%%%%%%%%%%%%%

In our paper \cite{Carrington:2021dvw} we calculated the coefficient $\hat{q}$ in Eq.~(\ref{tensor-X}) without the Wilson link operator that is required to make it gauge invariant. We argued that the gauge dependence of this quantity can be assessed by calculating the average value of the Wilson line itself. We therefore calculate 
\be
\Big\langle \! \big\langle W \big(t,\vec x \,|\, t',\vec x - \vec v (t-t')\big)  \big\rangle \! \Big\rangle  
\equiv \frac{1}{N_c^2-1}{\rm Tr}\big\langle W \big(t,\vec x \,|\, t',\vec x - \vec v (t-t')\big) \big\rangle \,,
\label{W-ave}
\ee
where the Wilson line is given by Eq.~(\ref{W-def}) and the trace is over the colour indices.

If the size of the quantity (\ref{W-ave}) is close to one for typical values of the arguments $t,t', \vec x$ and $\vec v$, it seems reasonable to claim that the gauge dependence introduced by setting the link operators to one is small. In this section we present a calculation of the average Wilson line (\ref{W-ave}) that is more accurate than the estimate we made in \cite{Carrington:2021dvw}. This calculation also serves as a warm-up for the much more difficult calculation of the gauge-independent momentum broadening coefficient. 

The calculation of the average in (\ref{W-ave}) involves two independent expansions: we expand the exponential of the Wilson line in powers of the potential $A$ and then we expand the potentials in powers of the proper time $\tau$. In our work  \cite{Carrington:2021dvw} we kept terms up to order $A^2$ and $\tau^0$ and obtained $\big\langle \! \langle W \rangle \! \big\rangle \approx 0.84$. 

In the MV limit the colliding nuclei are treated as infinite and uniform in the transverse plane and we can take the probe's velocity to lie along the $x$-axis, as already explained in Sec.~\ref{sec-corr}. With $\vec v = (v,0,0)$ the result for either $\big\langle \! \langle W \rangle \! \big\rangle$ or $\hat{q}$ should be uniform in the direction perpendicular to the probe's velocity. We therefore set these coordinates to zero after all derivatives have been taken. We additionally restrict the location of the hard probe to $z=0$ which means $t=\tau$. We also take into account that the time component of the potential given by the ansatz (\ref{ansatz}) vanishes for $z=0$. The path integral in the Wilson line (\ref{W-def}) can then be written as
\be 
\label{path-integral}
\int^{(t,\vec x)}_{(t',\,\vec y)} dz_\mu A^\mu_c(z) 
= - \int^{(t,x)}_{(t',\, y)} du \, \alpha^x_{\perp c} (t_u,u) 
= - v \int^t_{t'} dt_u \alpha^x_{\perp c} \big(t_u, v t_u +b \big),
\ee
where $t_u$ and $u$ are the time and $x$ components of the four-position argument of $A^\mu_c$ and $y \equiv x-v(t-t')$. The second equality holds under the assumption that the path integral is taken along the straight line from $y$ to $x$,  with $b \equiv x-vt$. The value of the intercept $b$ is irrelevant because our correlation functions are assumed translation invariant, as will be explained below.  
Using equation (\ref{path-integral}) and expanding the Wilson line in powers of the potential we have 
\ba
\nn
&& W(t, x |\, t', y)  = \mathds{1} 
- ig v \int^{t}_{t'} \, dt_u \,\alpha^x_{\perp a}(t_u, v t_u + b) T^a
\\[2mm]
\label{W-expansion}
&& ~
- g^2 v^2 \int_{t'}^{t} dt_u \, \alpha^x_{\perp a}(t_u, v t_u+b) T^a
\int_{t'}^{t_u} dt_w \, \alpha^x_{\perp b} (t_w, v t_w + b) T^b + \dots
\ea
where $\mathds{1}$ is an $(N_c^2-1) \times (N^2_c-1)$ unit matrix. We note that the third term satisfies the requirement of the path ordering of the potentials. 

We emphasize that in our calculation the small parameter is not the coupling constant $g$ but the proper time (actually $\tau Q_s$). Equation (\ref{W-expansion}) shows that each power of the potential is accompanied by an integral over an interval of order $t'$.  From Eqs.~(\ref{qhat-X-T}) and (\ref{tensor-X}) we know that $t'$ lies between zero and $t$, where $t$ is the time at which we calculate $\hat{q}$. The expansion of the Wilson line (\ref{W-expansion}) is justified by the fact that we consider only short times. 

We expand the gauge potentials in powers of the proper time (which coincides with time at $z=0$). The coefficients in the series (\ref{alpha-exp}) multiplying odd powers of time vanish. Omitting all arguments the even coefficients to fourth order in the fundamental representation are \cite{Carrington:2020ssh}
\be
\label{coefficients}
\begin{aligned}
& \alpha^{i(2)}_{\perp} = \frac{1}{4} \epsilon^{ij} [{\cal D}^j, B] ,
\\[2mm]
& \alpha^{i(4)}_{\perp} = \frac{ig}{64}  \big[ [{\cal D}^i,B], B \big] 
+ \frac{1}{64}\epsilon^{ij} \Big[{\cal D}^j, \big[{\cal D}^k, [{\cal D}^k, B ]\big]\Big] 
 + \frac{ig}{16}\big[\alpha^{(0)},[{\cal D}^i ,\alpha^{(0)}]\big] ,
\end{aligned}
\ee
where 
\be
\label{Dj-B}
{\cal D}^i \equiv \partial^i - ig \alpha_\perp^{i(0)},
~~~~~~~~
B \equiv \partial^y \alpha_\perp^{x(0)} - \partial^x \alpha_\perp^{y(0)} 
- ig [\alpha_\perp^{y(0)},\alpha_\perp^{x(0)}]
\ee
and $\epsilon^{ij}$ represents an antisymmetric matrix such that $\epsilon^{11}=\epsilon^{22}=0$ and $\epsilon^{12}=-\epsilon^{21}=1$. 

We substitute the potential $\alpha^x_{\perp a}$ expanded in the powers of time into Eq.~(\ref{W-expansion}), use the initial conditions to rewrite the resulting expression in terms of pre-collision potentials, and then average over the colour configurations of the initial nuclei and trace over the colour indices. The first term gives $\big\langle \! \langle \mathds{1} \rangle \! \big\rangle = {\rm Tr}[\langle \mathds{1} \rangle]/(N_c^2-1) = 1$. The trace of the second term vanishes because the generators $T^a$ are traceless. We also observe that due to the boundary condition (\ref{cond1}) and the formulas (\ref{coefficients}) and (\ref{Dj-B}), the terms proportional to odd powers of $g$ will give zero after using Wick's theorem because they are products of an odd number of pre-collision potentials. The first term that produces a non-trivial contribution to $\big\langle \! \langle W \rangle \! \big\rangle$ is the quadratic term denoted $W_2$ which is
\ba
\nn
&& W_2 = -g^2  v^2 \int_{t'}^{t} d t_u \int_{t'}^{t_u} d t_w \, 
\Big(\alpha^{x(0)}_{\perp a}( v t_u + b ) 
+ t_u^2\alpha^{x(2)}_{\perp a}(v t_u + b )+\cdots \Big)T^a
\\[2mm]
\label{W2-expanded}
&&~~~~~~~~\times
\Big(\alpha^{x(0)}_{\perp b}(v t_w + b )
+ t_w^2\alpha^{x(2)}_{\perp b}(v t_w + b ) +\cdots\Big) T^b.
\ea

We first consider the contribution to $\big\langle \! \langle W_2 \rangle \! \big\rangle$ from the zeroth order terms in the proper time expanded potentials, which is denoted $\big\langle \! \langle W_{20} \rangle \! \big\rangle$. We express $\alpha^{x(0)}$ in terms of the pre-collision potentials $\beta_1^x$ and $\beta_2^x$ using the boundary condition (\ref{cond1}) and take the expectation value. In the MV limit the correlators of pre-collision potentials are invariant under translations in the $x$-$y$ plane and thus  
\be
\label{corr-MV-xx}
\big\langle \beta_{n a}^x\big(v t_u +b \big) \beta_{n b}^x\big(v t_w +b\big) \big\rangle 
= \delta_{ab}B^{xx}\big(v(t_u-t_w)\big)~~~~ {\rm for}~~~~ n=1,2,
\ee 
where only non-vanishing position coordinates of the potentials $\beta_{n a}^i$ are written down explicitly and the operation $\lim_{{\rm w}\to 0}$ which is present in the correlator definition (\ref{field-corr}) is implicitly understood. 

We make two important observations about the expression in equation (\ref{corr-MV-xx}). The first is that the variable $b$ does not appear on the right side because of the translation invariance of the correlator of two pre-collision potentials. As explained above, $b\equiv x-vt$ is introduced so that the path integral is formally taken along a straight line path between the points $(t',y)$ and $(t,x)$, but due to translation invariance the value of this intercept plays no role. The second point is that since $t_u>t_w$ we do not need absolute values on the argument of the correlation function on the right side. We emphasize that while it is true that in the lowest order calculation we are currently discussing there is no difference between $\int_{t'}^{t_u} dt_w B^{xx}(v(t_u-t_w))$ and $\int_{t'}^{t_u} dt_w B^{xx}(v|t_u-t_w|)$, it is important that the absolute values not be used at higher orders. This is because at higher orders we obtain derivatives of correlation functions and if we write $B^{xx}\big(v|t_u - t_w|\big)$ then there is an ambiguity about the sign of $B^{xx}_{(0,1)}\big(v|t_u - t_w|\big)$. 

From equations (\ref{W2-expanded}) and (\ref{corr-MV-xx}) we obtain 
%\be
%\label{my20-old}
%\big\langle \! \langle W_{20} \rangle \! \big\rangle
%=  - 2 g^2\,N_c\, v^2\int_{t-t'}^{t} dt_u\int_{t-t'}^{t_u} dt_w \, B^{xx}\big(v(t_u-t_w)\big) ,
%\ee
\be
\label{my20}
\big\langle \! \langle W_{20} \rangle \! \big\rangle
=  - 2 g^2\,N_c\, v^2\int_{t'}^{t} dt_u\int_{t'}^{t_u} dt_w \, B^{xx}\big(v(t_u-t_w)\big) ,
\ee
where the identity ${\rm Tr}[T^a T^a] = N_c(N_c^2-1)$ has been used. The factor 2 comes from the sum of two correlators of pre-collision potentials from each of the two nuclei. 

The contribution $\big\langle \! \langle W_{20} \rangle \! \big\rangle$ will be counted as second order in time, in spite of the fact that it is obtained from expanded potentials at zeroth order, because of the two time integrals which are each over an interval of order $t$. To understand this we note that replacing the correlator in Eq.~(\ref{my20}) by a constant and computing the double time integral one finds that $\big\langle \! \langle W_{20} \rangle \! \big\rangle$ is proportional to $(t-t')^2$. Since $0 \le t' \le t$ it is clear that the integral is of order $t^2$.

Next we calculate the contribution to $\big\langle \! \langle W_2 \rangle \! \big\rangle$ from second order terms in the proper time expanded potentials, which is denoted $\big\langle \! \langle W_{22} \rangle \! \big\rangle$. Proceeding as in case of $\big\langle \! \langle W_{20} \rangle \! \big\rangle$, a straightforward calculation gives 
\ba
\nn
\big\langle \! \langle W_{22} \rangle \! \big\rangle 
&=& -\frac{1}{2}g^2\, N_c v^2 \int_{t'}^{t} d t_u  \int_{t'}^{t_u} dt_w \,
(t_u^2+t_w^2) \Big[ \Big( B^{xx}_{(2,0)}\big(v(t_u - t_w)\big) - B^{xx}_{(1,1)}\big(v(t_u - t_w)\big)\Big) 
\\ [2mm] 
\label{my-ord2}
&& ~~~~~~~~~~
- 2g^2 N_c \Big( B^{xx}\big(v(t_u - t_w)\big) \, B^{yy}(0) - B^{xy}\big(v(t_u - t_w)\big) \, B^{xy}(0) \Big) \Big] .
\ea

The result for $\big\langle \! \langle W_{22} \rangle \! \big\rangle$ in (\ref{my-ord2}) is of order $t^4$, but it is not the only contribution to $\big\langle \! \langle W \rangle \! \big\rangle$ at fourth order. An additional fourth order contribution is obtained when the path-ordered exponential is expanded to fourth order and the potentials are calculated at zeroth order in the proper time expansion. This contribution is denoted $\big\langle \! \langle W_{40} \rangle \! \big\rangle$ and  is computed similarly to $\big\langle \! \langle W_{20} \rangle \! \big\rangle$. The complete result up to fourth order is 
\be
\big\langle \! \langle W \rangle \! \big\rangle_4 = 1 + \big\langle \! \langle W_{20} \rangle \! \big\rangle + \big\langle \! \langle W_{22} \rangle \! \big\rangle + \big\langle \! \langle W_{40} \rangle \! \big\rangle .
\ee
The correlation functions in equations (\ref{my20}) and (\ref{my-ord2}) are given in equation (\ref{deriv-res}) in terms of the functions $C_1 (r)$ and $C_2(r)$ which are defined in (\ref{C1-def}), (\ref{C2-def}), and it is straightforward to obtain similar expressions for the correlators that are needed to calculate $\big\langle \! \langle W_{40} \rangle \! \big\rangle$.  We remind the reader that the function $C_1(r)$ is logarithmically divergent as $r \to 0$ and the method we used to regulate it is explained in detail in Sec.~\ref{sec-regular}. 

We have computed  $\big\langle \! \langle W \rangle \! \big\rangle_4$ and $\big\langle \! \langle W \rangle \! \big\rangle_2$ for a hard probe moving at the speed of light perpendicularly to the beam axis at $z=0$. For comparison we also consider the incomplete fourth order result $\big\langle \! \langle W \rangle \! \big\rangle_{2+}$ that does not include the contribution from $\big\langle \! \langle W_{40} \rangle \! \big\rangle$, which is defined as $\big\langle \! \langle W \rangle \! \big\rangle_{2+} = 1 + \big\langle \! \langle W_{20} \rangle \! \big\rangle + \big\langle \! \langle W_{22} \rangle \! \big\rangle$. The coupling constant $g$, saturation scale $Q_s$, and infrared cutoff $m$ are chosen to be  $g=1$, $Q_s = 2$ GeV and $m=0.2$ GeV, and we take $N_c=3$. 

\begin{figure}[t]
\begin{center}
\includegraphics[width=11cm]{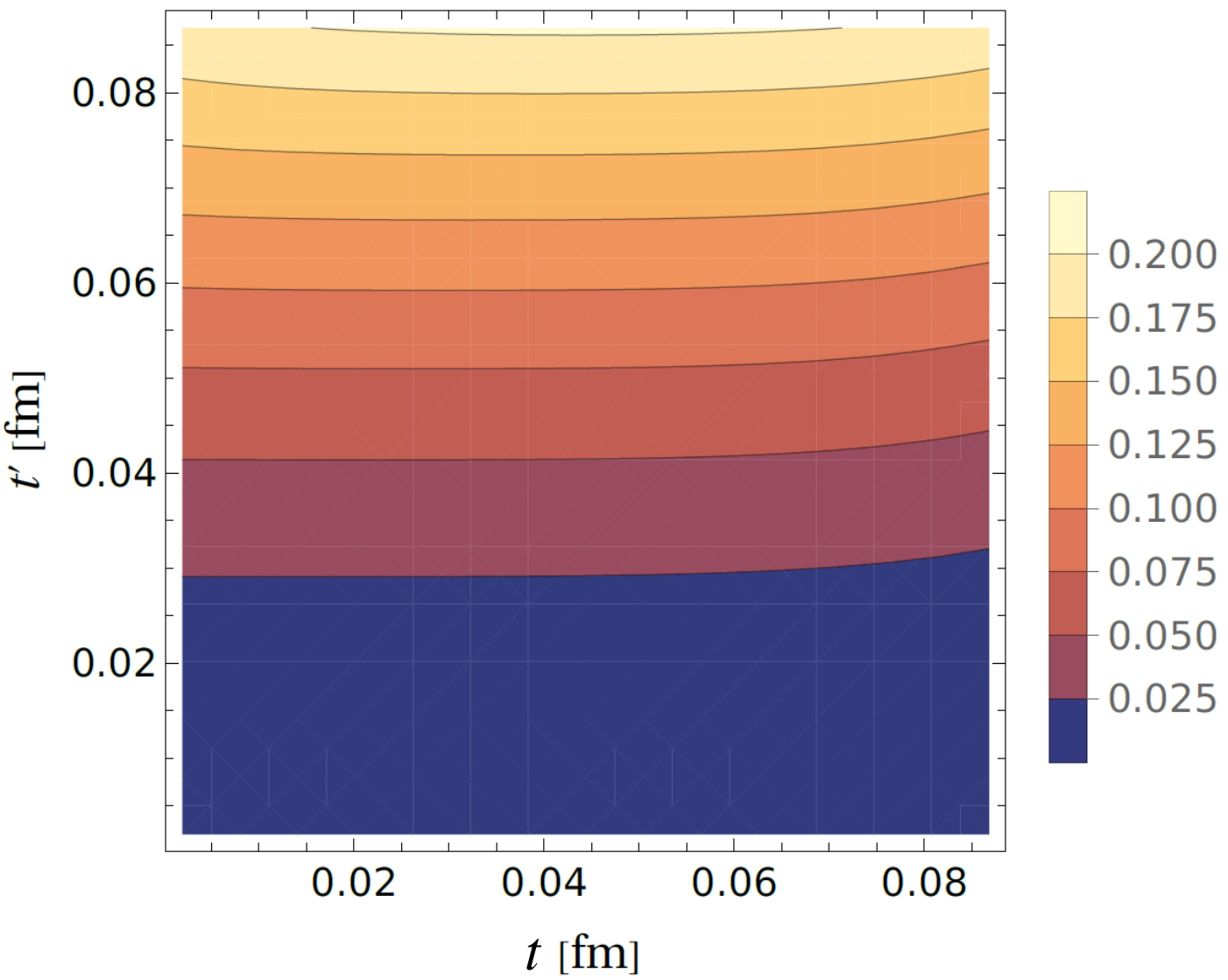}
\end{center}
\vspace{-7mm}
\caption{A contour plot of $1 - \big\langle \! \langle W(t,t-t') \rangle \! \big\rangle_4$} as a function of $t$ and $t'$ both in fm.
\label{plot-Walone-1}
\end{figure}

\begin{figure}[t]
\begin{center}
\includegraphics[width=11cm]{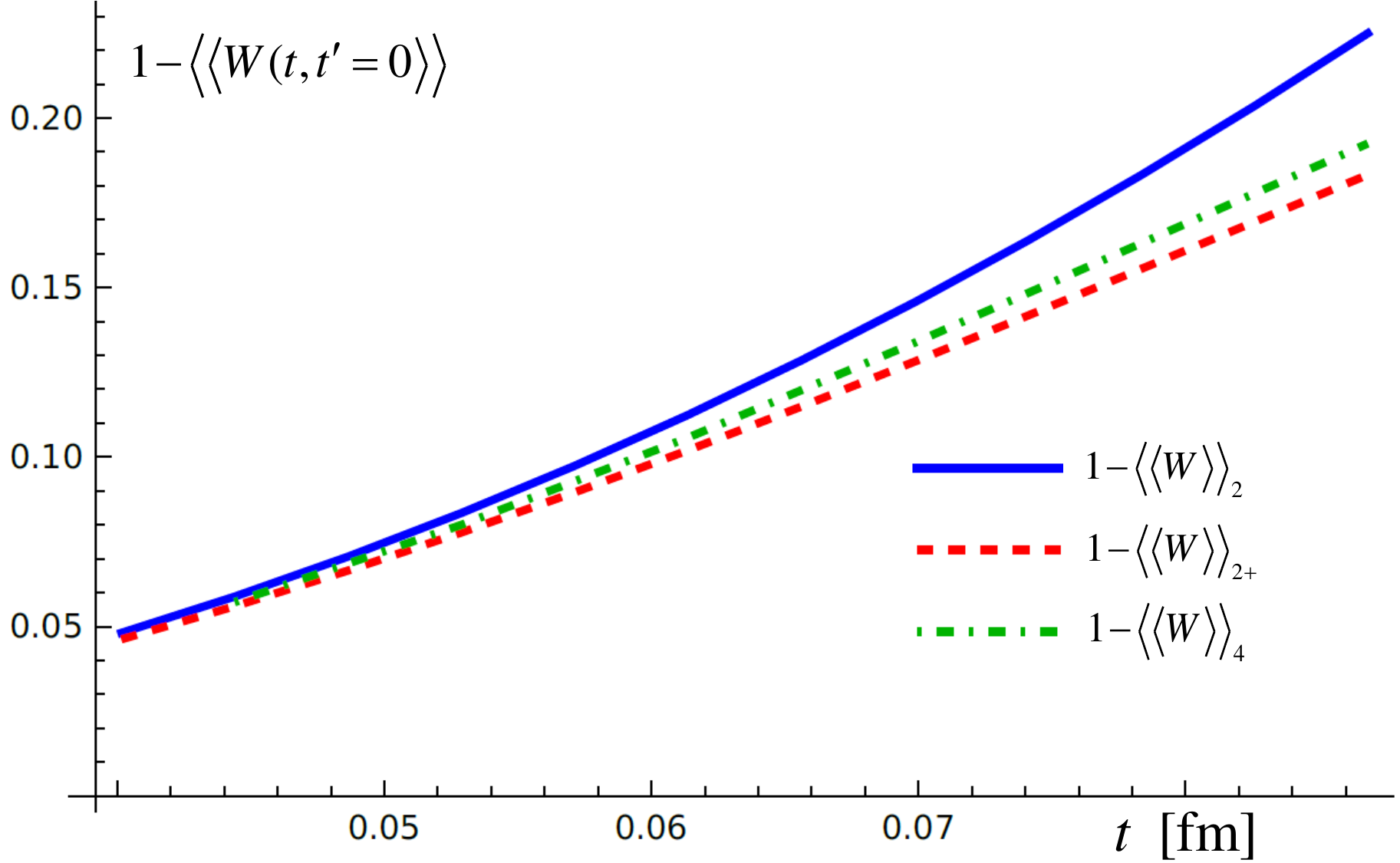}
\end{center}
\vspace{-7mm}
\caption{$1- \big\langle \! \langle W(t,t'=0) \rangle \! \big\rangle$ as a function of $t$ in fm at the second, fourth and incomplete fourth orders of the proper time expansion (see text for further explanation).}
\label{plot-Walone-2}
\end{figure}

Fig.~\ref{plot-Walone-1} shows a contour plot of $1 - \big\langle \! \langle W \rangle \! \big\rangle_4 $ versus $t'$ and $t$. The dependence on $t$ is very weak because the only parts of the integrand that are not fully translationally invariant in time are the factors that multiply the correlators. In  Fig.~\ref{plot-Walone-2} we show the result to second order, to fourth order, and the partial fourth order result, with $t=t'$. 

Figures \ref{plot-Walone-1} and \ref{plot-Walone-2} show that $\big\langle \! \langle W \rangle \! \big\rangle$ is close to one over the whole range where the proper time expansion is valid, which is $t \lesssim 0.06$~fm. Based on these results  one would estimate that a gauge invariant calculation of $\hat q$ would differ from our previous result by 10-15\%. However, this estimation should be treated only as a rough approximation, since it is obtained by factoring out the Wilson line in Eq.~(\ref{tensor-X}). This factorization can be written schematically as $\langle F_a W_{ab} F_b \rangle \to \langle F_a F_b\rangle\langle W_{ab} \rangle$. An additional issue is that our results for $\big\langle \! \langle W \rangle \! \big\rangle$ include terms with correlators of pre-collision potentials from only one ion. These terms represent the initial nuclei but not the glasma, and consequently were ignored in our calculations of $\hat{q}$ in \cite{Carrington:2022bnv,Carrington:2021dvw,Carrington:2023nty}. When we calculate the expectation value of the Wilson line by itself, correlators of potentials from the same ion must be kept, to consistently calculate corrections to the leading order contribution: $\big\langle \! \langle W \rangle \! \big\rangle = 1+ \dots$. To resolve these issues we compute the gauge invariant coefficient $\hat{q}$ in the next section.

%%%%%%%%%%%%%%%%%%%%%%%%%%%%%%%%%%%%%%%%%%%%%%%%%%%%%%
\section{Gauge invariant momentum broadening}
\label{sec-qhat}
%%%%%%%%%%%%%%%%%%%%%%%%%%%%%%%%%%%%%%%%%%%%%%%%%%%%%%

The computation of the momentum broadening coefficient $\hat{q}$ in Eqs.~(\ref{qhat-X-T}) and (\ref{tensor-X}) with $W = \mathds{1}$ is described in detail in our earlier paper \cite{Carrington:2022bnv} where the results up the fifth order in the proper time expansion are given, and in \cite{Carrington:2023nty} we extended these calculations to seventh order. The corresponding gauge invariant calculation is much more difficult because not only the fields explicitly present in the tensor (\ref{tensor-X}) must be expanded in powers of the proper time, but the Wilson line must also be expanded. The number of terms to be integrated is dramatically bigger and more importantly the integrals that must be calculated have a much more complicated structure. For technical reasons it is easier to do the calculation in the fundamental representation, because the colour matrices have lower dimension. We write the Wilson line (\ref{W-def}) as
\be
W_{ab}(t,\vec x \,|\, t',\vec y) = 
2 {\rm Tr}[ t^a P(t,\vec x \,|\, t',\vec y) t^b P^\dagger(t,\vec x \,|\, t',\vec y) ] ,
\ee
where $P(t,\vec x \,|\, t',\vec y) $ is the Wilson line (\ref{W-def}) with the adjoint matrix $T^c$ replaced by fundamental generator $t^c$.

We have obtained gauge invariant results for $\hat{q}$ only up to cumulative fourth order in the proper time expansion. It should be understood that our results are invariant up to contributions from higher order terms, since the Wilson line is expanded in the proper time. The issue of gauge invariance is discussed in detail at the end of this section.

\begin{figure}[t]
\begin{center}
\includegraphics[width=11cm]{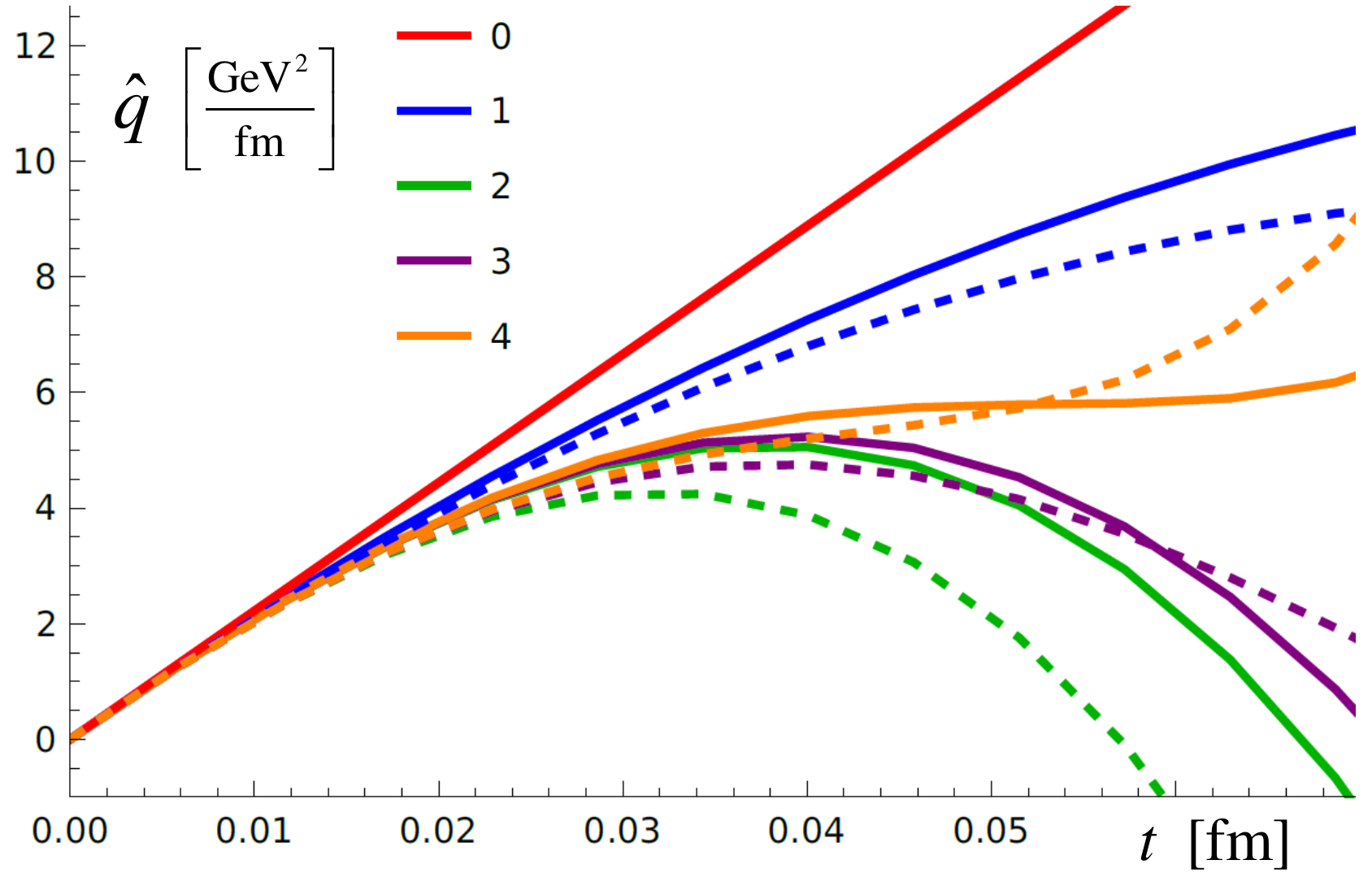}
\end{center}
\vspace{-8mm}
\caption{The momentum broadening coefficient $\hat q$ in GeV$^2$/fm versus time in fm. The solid lines show the result obtained in \cite{Carrington:2022bnv} with $W = \mathds{1}$ and the dashed lines are the results from the gauge invariant calculation.}
\label{plot-FWF}
\end{figure}

\begin{figure}[t]
\begin{center}
\includegraphics[width=11cm]{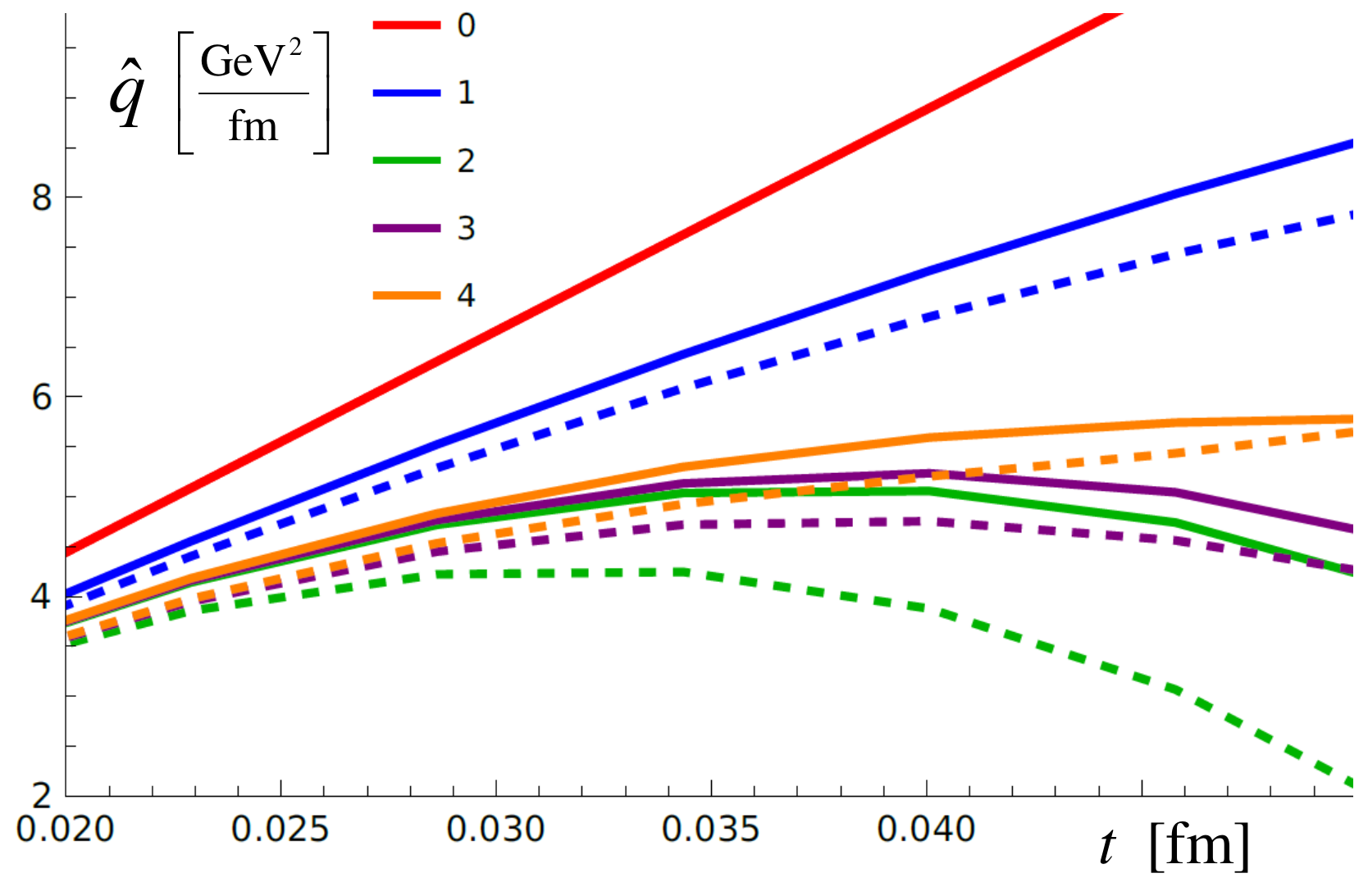}
\end{center}
\vspace{-8mm}
\caption{A close-up of the central region in Fig.~\ref{plot-FWF}.}
\label{plot-FWF-sat}
\end{figure}

We show our results for $\hat{q}$ in Fig.~\ref{plot-FWF} where our previous results obtained using the approximation $W = \mathds{1}$ are represented by the solid lines and the gauge invariant results are shown as dashed lines. As in Sec.~\ref{sec-Wilson-line}, the hard probe moves at the speed of light perpendicularly to the beam axis at $z=0$ and we take $N_c = 3$, $g=1$, $Q_s = 2$ GeV and $m=0.2$ GeV. The calculations are performed at cumulative fourth order in $t$. Figure~\ref{plot-FWF-sat} is a close-up of the region where $\hat{q}$ reaches its maximal value and the proper time expansion is still reliable. The graphs demonstrate that including the Wilson line does not strongly affect the time dependence of $\hat{q}$, since the shape, height and location of the maximum are largely unchanged. At zeroth order all results are trivially identical and the influence of the Wilson line initially grows as the order of the expansion increases and then declines. 

We note that the prominent rise towards the upper part of the $t$ range of the dashed orange curve in Fig.~\ref{plot-FWF} is not a physical effect. This simply signals a breakdown of the proper time expansion performed up to 4th order. Our calculations that do not include the Wilson lines show that our results for $\hat{q}$ are reliable at longer times $t \lesssim 0.06~{\rm fm}/c$ when contributions up to 8th are taken into account \cite{Carrington:2023nty}.

We want to quantitatively compare our  results for $\hat{q}$ at fourth order which are shown in Figs.~\ref{plot-FWF} and \ref{plot-FWF-sat}. It would be natural to compare the maximal values of $\hat{q}$, but this does not work because there is no global maximum at fourth order. This happens because at fourth order $\hat q(t)$ diverges towards positive infinity when the proper time expansion breaks down. We therefore compare the values of $\hat q$ at the inflection point where the second derivative of $\hat q(t)$ is zero. This gives $\hat q_{\rm infl}=5.79~{\rm GeV^2/fm}$ from the calculation with $W = \mathds{1}$ and $\hat q_{\rm infl}=5.28~{\rm GeV^2/fm}$ from the gauge invariant calculation. These values show that at fourth order the gauge invariant result for the momentum broadening coefficient differs by approximately 9\% from the result obtained with the Wilson line set to unity. 

There is a subtle issue regarding the gauge invariance of our results that should be considered carefully. We remind the reader that our calculation involves two independent expansions: the proper time expansion that is used to solve the YM equations and the expansion of the Wilson line. Our method for combining these two expansions is explained under equation (\ref{my20}). The basic idea is that although the exponential in the Wilson line is formally expanded in $g$, the coupling constant is not the relevant small parameter. Each power of the coupling multiplies an integral with length proportional to the proper time. In our counting of orders we take each integral from an expanded Wilson line to be of order $\tau^1$ and combine this with the powers of $\tau$ obtained from the proper time expansion of the gauge potentials. This means that, for example, the fourth order result includes some but not all of the contributions from the $g^4$ term in the expansion of the Wilson line. The terms that are dropped are fifth order in $\tau$, using our counting procedure, and they are significantly smaller than the terms that are kept. The consequence however is that while our calculation is gauge invariant to the order at which we work, it is not formally correct to say that it is fully gauge invariant. Our numerical results in Figs.~\ref{plot-FWF} and \ref{plot-FWF-sat} indicate that the effect of the terms in the expanded Wilson line that are not included is negligible. 

\begin{figure}[t]
\begin{center}
\includegraphics[width=12cm]{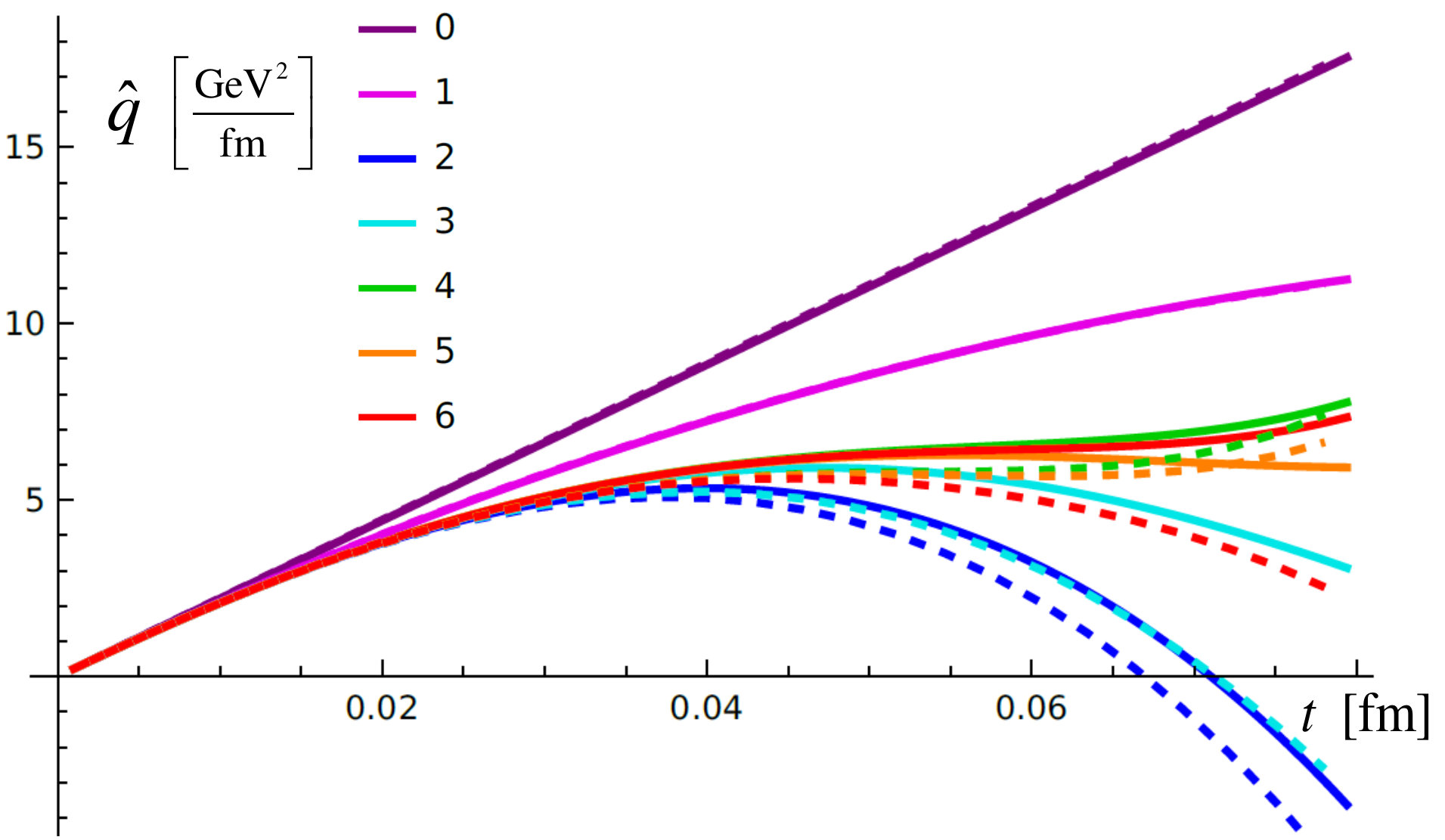}
\end{center}
\vspace{-7mm}
\caption{The momentum broadening coefficient $\hat q$ in GeV$^2$/fm versus time in fm. The coefficient is computed with $W = \mathds{1}$, using the original regularization method (solid lines) and the new method (dashed lines). The dashed pink line is not visible because it is directly under the solid pink line.}
\label{untwist}
\end{figure}

Finally, we return to the issue mentioned at the end of Sec.~\ref{sec-corr}. For technical reasons we have used a slightly different method to regulate the ultraviolet divergences in the two-point correlator, relative to our previous calculations. Specifically, the limit in Eq.~(\ref{one-point-def}) that changes a one-point correlator into a two-point correlator can be ill defined for path ordered products of potentials from the expanded Wilson line. A comparison of the results obtained from the two regularizations with $W = \mathds{1}$ is shown in Fig.~\ref{untwist}. The calculation is done using the same parameters as previously. The graph shows that although the results are not identical, they are very similar. This indicates that our results are largely independent of the regularization method.

%%%%%%%%%%%%%%%%%%%%%%%%%%%%%%%%%%%%%%%%%%%%%%%%%%%%%%
\section{Conclusions}
%%%%%%%%%%%%%%%%%%%%%%%%%%%%%%%%%%%%%%%%%%%%%%%%%%%%%%

Our calculations of the momentum broadening coefficient $\hat{q}$ presented in \cite{Carrington:2022bnv,Carrington:2021dvw,Carrington:2023nty} did not satisfy the requirement of gauge independence. We gave only an estimate of the possible effect of this shortcoming based on the size of a colour averaged Wilson line. In this work we have done a more accurate calculation of a colour averaged Wilson line and confirmed that this average is approximately one. The main result of this paper is our calculation of the momentum broadening coefficient using a gauge invariant formulation.  The results differ by about 9\% from our previous results which were obtained using a simplifying approximation that violated gauge invariance. The calculation presented in this paper therefore confirms our previous conclusion about the significant role of the glasma in jet quenching. 

%-----------------------------------------------------------------------
\section*{Acknowledgments}
%-----------------------------------------------------------------------

This work was partially supported by the Natural Sciences and Engineering Research Council of Canada under grant SAPIN-2023-00023. 

%-----------------------------------------------------------------------
\section*{Data availability}
The data that support the findings of this article are openly
available \cite{dataWL}.
%-----------------------------------------------------------------------

\appendix

%%%%%%%%%%%%%%%%%%%%%%%%%%%%%%%%%%%%%%%%%%%%%%%%%%%%%%
\section{Derivation of formula for $\hat{q}$}
\label{app-qhat-deriv}
%%%%%%%%%%%%%%%%%%%%%%%%%%%%%%%%%%%%%%%%%%%%%%%%%%%%%%

The coefficient $\hat q$ that measures the momentum broadening per unit length of a test parton in the direction transverse to its initial velocity $\vec v$ is defined as
\ba
\label{qq}
\hat q &=& \frac{1}{v} \Big( \delta^{\alpha\beta} -\frac{v^\alpha v^\beta}{v^2} \Big)
\lim_{\Delta t \to 0} \frac{\langle \Delta p^\alpha \Delta p^\beta \rangle}{\Delta t} \,,
\ea
where $\Delta p^\alpha$ is the momentum change of the parton in the time interval $\Delta t$ and $\langle \dots \rangle$ denotes averaging over an ensemble of test partons. To calculate the correlator on the right side of (\ref{qq}) we start with the equation of motion for a hard probe which is the Wong equation
\be
\label{wong2}
\frac{d\vec{p}(t)}{dt} =q^a \vec{F}_a\big(t,\vec{r}(t)\big) ,
\ee
where $\vec p(t)$ and $\vec r(t)$ are the momentum and trajectory of a hard probe, and the chromodynamic Lorentz force is $\vec F_a\big(t,\vec{r}(t)\big) \equiv g  \Big(\vec{E}_a\big(t,\vec{r} (t)\big)  +\vec{v}(t)\times\vec{B}_a\big(t,\vec{r} (t)\big)  \Big)$ with $q^a$ the colour charge of the hard probe. The momentum change of the probe during the time interval from 0 to $t$ is
\be
\Delta p^\alpha = q^a\int_0^t dt' F_a^\alpha\big(t',\vec{r}(t')\big) 
\ee
and therefore 
\be
\label{Delta-pi-Delta-pj-QCD}
\langle \Delta p^\alpha \Delta p^\beta \rangle = \int_0^t dt'  \int_0^t dt'' \frac{1}{N_{cs}}\int Dq \, q^a q^b
\big\langle F^\alpha_a\big(t',\vec{r}(t')\big) F^\beta_b\big(t'',\vec{r}(t'')\big) \big\rangle,
\ee
where the angle brackets indicate both averaging over an ensemble of colour sources and averaging the colour charge of the probe. We use the notation $N_{cs}$ for the number of colour states of the hard probe which is $N_c$ for a quark and $N_c^2-1$ for a gluon. Since averaging over parton colours gives \cite{Litim:2001db} 
\be
\label{int-qq}
\int Dq \,q^a q^b = \delta^{ab}  \left\{
\begin{array}{ccl}
\frac{1}{2} &{\rm for} & {\rm a~quark} , 
\\[3mm]
 N_c &{\rm for}  & {\rm a~gluon} ,
\end{array}
\right. 
\ee
equation~(\ref{Delta-pi-Delta-pj-QCD}) becomes 
\be
\label{Delta-pi-Delta-pj-QCD-2}
\langle \Delta p^\alpha \Delta p^\beta \rangle = C_R \int_0^t dt'  \int_0^t dt''
\big\langle F^\alpha_a\big(t',\vec{r}(t')\big) F^\beta_a\big(t'',\vec{r}(t'')\big) \big\rangle 
\ee
with the colour factor $C_R$ equal to $1/(2N_c)$ for a quark and $N_c/(N_c^2-1)$  for a gluon. 
Assuming only that the integrand in (\ref{Delta-pi-Delta-pj-QCD-2}) is a smooth function we can write 
\ba
\nn
\lim_{\Delta t \to 0}\frac{\Delta p^\alpha \Delta p^\beta}{\Delta t} 
&=& C_R \frac{d}{dt} \int_0^t dt'  \int_0^t dt''
\big\langle F^\alpha_a\big(t',\vec{r}(t')\big) F^\beta_a\big(t'',\vec{r}(t'')\big) \big\rangle  
\\ \nn
&=& C_R \int_0^t dt'  \big\langle  F_a^\alpha\big(t,\vec{r}(t)\big) F_a^\beta\big(t',\vec{r}(t')\big) 
+  F_a^\alpha\big(t',\vec{r}(t')\big) F_a^\beta\big(t,\vec{r}(t)\big) \big\rangle 
\\
&\equiv& X_{\alpha\beta}+X_{\beta\alpha}  \, .\label{DpDp}
\ea
Substituting (\ref{DpDp}) into (\ref{qq}) we obtain 
\bea
\label{qhatfromX}
\hat q = \frac{1}{v} \Big( \delta^{\alpha\beta} -\frac{v^\alpha v^\beta}{v^2} \Big)(X_{\alpha\beta}+X_{\beta\alpha})
= \frac{2}{v} \Big( \delta^{\alpha\beta} -\frac{v^\alpha v^\beta}{v^2} \Big)X_{\alpha\beta}
\eea
with
\bea
\label{Xij-QCD-def}
 X^{\alpha\beta} \equiv C_R \int_0^t dt'  \Big\langle  F^\alpha_a\big(t,\vec{r}(t)\big) F^\beta_a\big(t',\vec r(t') \big) \Big\rangle .
\eea
For quarks the result in (\ref{qhatfromX}) and (\ref{Xij-QCD-def}) agrees with equations (\ref{qhat-X-T}) and (\ref{tensor-X}) with $W = \mathds{1}$, which is the formula obtained in \cite{Mrowczynski:2017kso} using a Fokker-Planck approach.

%\newpage

\end{document}